\documentclass{iau-JDSS}
\usepackage{graphicx}

\title[JD 03.~~Stellar Activity Cycles] 
{Theoretical Models of Stellar Activity Cycles}

\author[E. I\c{s}\i k]   
{Emre I\c{s}\i k}

\affiliation{Department of Physics, Istanbul K\"ult\"ur University,
Bak\i rk\"oy, 34156, Istanbul, Turkey \break e-mail: e.isik@iku.edu.tr}

\pubyear{2013}
\volume{Volume 16}  
\pagerange{119--120}
\date{?? and in revised form ??}
\jname{Highlights of Astronomy, Volume 16}
\editors{Thierry Montmerle, ed.}
\begin{document}

\maketitle

\begin{abstract}
We discuss possible mechanisms underlying the observed features of stellar activity cycles, such as multiple periodicities in very active stars, non-cyclic activity observed in moderately active stars, and spatial distribution of stellar magnetic regions. We review selected attempts to model the dependence of stellar activity cycles on stellar properties, and their comparison with observations. We suggest that combined effects of dynamo action, flux emergence and surface flux transport have substantial effects on the long-term manifestations of stellar magnetism.  
\end{abstract}

\firstsection 
\section{Modelling stellar magnetic cycles}

Cool stars with convective envelopes exhibit long-term photometric and 
spectroscopic variations, which indicate long-term changes of 
surface magnetic flux. Many stars show single or multiple cycles with some 
irregularities, while others are 
characterised by relatively unchanging total radiative flux, hence they 
are called \emph{flat activity stars}. Whether these stars represent  
grand minimum states is not yet certain. 

The current paradigm concerning the physics of activity cycles in cool stars 
is based on solar dynamo models, which incorporate the generation and 
transport of magnetic flux within the convection zone. 
There are several unknowns regarding rapidly rotating cool stars, in particular 
the internal rotation, flows, magnetic field geometry, and their complex interplay. Therefore it is probably more appropriate to extrapolate the solar 
paradigm \emph{gradually} towards faster rotating Sun-like stars or to cooler dwarfs. 
We expect the dynamo strength to increase with decreasing Rossby number, 
in parallel with the empirical rotation-activity relation. 
Simple mean-field dynamo models 
suggest that the cycle period decreases as the dynamo number increases. 
However, recent flux transport dynamo 
models by \cite{jouve10} have shown that, to match the observed dependence, 
meridional flow should either increase its strength with the rotation rate 
(hereafter $\Omega$), which is incompatible with 
3D full-sphere simulations, or have preferred patterns which are difficult to justify. 
As an alternative explanation 
for the observed relation, \cite{dbrun12} have shown that turbulent 
pumping of magnetic flux throughout the convection zone can 
decrease the cycle period with increasing $\Omega$ and decreasing meridional 
flow speed. However, it should be noted that recent mean-field models of 
differential rotation 
indicate an increase of meridional flow speed with $\Omega$ 
(\cite[Kitchatinov \& Olemskoy 2012]{ko12}). 
Magnetoconvection simulations in 3D of the 
entire stellar convection zone have been presented by \cite{brown11}, who 
demonstrated self-organisation of large-scale toroidal magnetic structures in the 
midst of the convection zone of a Sun-like star. 
These structures turn out to be anti-symmetric around 
the equator, stationary for 3$\Omega_\odot$, and 
exhibit occasional polarity reversals for $5\Omega_\odot$. 

We have recently presented numerical simulations combining generation, 
buoyant rise, and surface transport of magnetic flux in various cool star 
convection zone configurations (\cite{isik11}). 
Our composite models are based on a mean-field overshoot dynamo, 
the period of which
decreases with $\Omega$. With increasing $\Omega$ the latitudinal distribution 
of the emerging flux at the surface deviates significantly from that of the dynamo 
waves at the base of the convection zone, owing to a poleward deflection of 
rising flux tubes. 

When the rotation period is 9~days for a Sun-like star, our models indicate that 
the periodic dynamo in the deep convection zone can lead to a non-cyclic 
surface activity. The existence of such moderately active 
but non-cycling stars were reported by \cite[Hall \& Lockwood (2004)]{hl04}. 
A similar situation can occur 
for coronal X-ray cycles in some cool stars: \cite{mcivor06} have shown that 
sufficiently overlapping magnetic cycles in the surface activity can lead to the 
absence of X-ray cycles. 

Considering mean-field dynamo models, \cite{moss11} have suggested that the 
dynamo must be operating in two distinct layers within the convection zone, 
to explain the observed 
poleward moving starspots in the K1-subgiant component of HR 1099. 
Our simulations of a K1-type subgiant star 
with $\Omega$ and 
surface shear adopted from HR 1099 have demonstrated that a deep-seated stellar 
dynamo with a well-defined period and amplitude can co-exist with fluctuating 
cycles of surface magnetic flux (\cite[I\c{s}\i k et al. 2011]{isik11}). 
Further analysis indicates that the short-period 
cycle at the surface results from the 
underlying dynamo, while the long-period cycle signal is caused by stochastic 
emergence of bipolar regions at very high latitudes, where 
the effects of surface transport are relatively weak (\cite{isik12}).
This can be an explanation for 
the long-term 'cycle' in HR 1099 reported by \cite{olah09}. 

\section{Open problems and outlook}

It is clear that we still lack a rigorous physical understanding of the relations 
between the emerging flux, rotation rate, convection zone structure, and cycle 
properties of cool stars (\cite{rempel08}). 
How does the link between a tachocline dynamo and surface transport 
change with rotation rate and convection zone structure, as we depart from solar 
parameters?
How do turbulent transport properties change with $\Omega$?
How does the dynamical disconnection of 
rising magnetic flux tubes vary with stellar parameters? 

Such problems will remain as challenges in the future, but observations 
will reduce the theoretical degrees of freedom. 
The recent
successes of the Babcock-Leighton type solar dynamo models in explaining various 
aspects of solar magnetism is encouraging for stellar modelling. 
Based on our recent models, we conclude that the combined effects of 
flux generation, emergence and transport should be considered in models 
of stellar magnetic fields, because their 
interrelations are sensitive to changes in stellar parameters.

\end{document}